\documentclass[]{aa}
\usepackage{amsmath}
\usepackage{amsfonts}

\newcommand{\be}{\begin{equation}}
\newcommand{\ee}{\end{equation}}

\newcommand{\bean}{\begin{eqnarray}}          
\newcommand{\eean}{\end{eqnarray}}

\newcommand{\bea}{\begin{eqnarray*}}          
\newcommand{\eea}{\end{eqnarray*}}

\def\Sum{\sum\limits}

\newcommand{\mx}[1]{\mbox{#1}}

\def\B{\mathcal{B}}

\def\v{\mx{v}}
\def\x{\mx{x}}

\def\Eq#1{(\ref{#1})}

\begin{document}

\title{Collective relaxation of stellar systems revisited}
\author{V.G.~Gurzadyan\inst{1},  A.A.~Kocharyan\inst{1,2}}

\institute
{\inst{1} Yerevan Physics Institute and Yerevan State University, Yerevan,
Armenia\\
\inst{2} School of Mathematical Sciences, Monash University, Clayton, Australia
}

\date{Received (\today)}

\titlerunning{Relaxation}

\authorrunning{V.G.Gurzadyan and A.A.Kocharyan}

\abstract{The chaos in stellar systems is studied using the 
theory of dynamical systems and the Van Kampen stochastic differential equation approach.
The exponential instability (chaos) of spherical N-body gravitating systems, already known previously, is confirmed.
The characteristic timescale of that instability is estimated confirming the collective relaxation time 
obtained by means of the Maupertuis principle.
}


\maketitle

\section{Introduction}

Stellar dynamics, a well established discipline, continues to be an active field of research 
due to its complex dynamics. Newly developed sophisticated tools on nonlinear dynamics or ever increasing numerical capacities are readily applied to N-body gravitating systems.

The problem of relaxation of gravitating systems is among the key ones in stellar dynamics since it can be directly constrained by observations, especially, of well-relaxed globular clusters and elliptical galaxies. 

Historically, plasma methods were among the first to be applied to gravitating systems (\cite{Chandra}),  neglecting, however, a drastic difference between plasma and long-range gravity. In retrospect, it can seem strange how easily were ignored the challenges in theory and observations, and later also in numerical simulations: (a) the Coulomb logarithmic cut-off and hence the canceling of the N-body effects was applied at the absence of Debye screening; (b) the result contradicted observations, i.e. the plasma two-body relaxation timescale  exceeds the age of elliptical galaxies by several orders of magnitude; (c) the two-body timescale could never be identified in numerical studies.

Then comes the epoch of realization of importance of chaos, occasionally in unexpected forms, for nonlinear systems. 
Chaos caused by small perturbations appears crucial even to nearly integrable problems, such as the dynamics of planetary system, (see \cite{Laskar,Morb} and refs therein); for non-integrable N-body systems the situation is far more complex. The Fermi-Pasta-Ulam problem is another example of apparently simple, but still not well understood system (\cite{FPU}).    

Do chaos, chance and randomness have significant role also in the evolution of stellar systems, or the two-body plasma approach is the whole story? Although it has been generally agreed that chaos must affect N-body dynamics (see \cite{GP,Cont,Reg}), proper treatments require well-founded approaches\footnote{Ruelle mentions the appearance of numerous incorrect papers about chaos when chaos became a fashion (\cite{R}).}. Application of ergodic theory methods enabled us to prove a notable result: the spherical systems are exponentially unstable and that chaoticity drives their dynamics (\cite{GS}). An exponential law defines an intrinsic timescale, the collective relaxation time, which for real stellar systems has a value intermediate between the dynamical (crossing) and the two-body timescales. This was obtained by estimating the divergence of the trajectories of the system in a Riemannian space defined by the potential of the interaction (Maupertuis principle), i.e. by a method known in theory of dynamical systems (\cite{Arnold}).
Importantly, the derived collective (N-body) relaxation timescale fits the observational data (\cite{Vesp}). The three timescales, i.e., the dynamical $\tau_{dyn}$, collective $\tau_{cr}$, and two-body $\tau_{b}$ timescales, correspond to the 3 distance scales of the system, specifically, its size $D$, mean inter-particle distance $d$, and radius of gravitational influence of particles $r_h$, in particular (\cite{GS}) (see also \cite{Lang}), 
\begin{equation}
\label{RelTime}
\tau_{cr}\sim \frac{D}{d} \tau_{dyn}\sim\tau_{dyn}N^{1/3}.
\end{equation} 
When the role of complex N-body dynamics was finally recognized, another confusion did appear, namely, in assigning of the dynamical time as the relaxation timescale, i.e., for reaching fine-grained equilibrium. However, this again contradicts observations: globular clusters would then have already disappeared because of the evaporation of stars (\cite{Amb}) within 100 crossing times, i.e., within around 100 mln years.\footnote{One of the motivations for concluding that the dynamical time equals the relaxation time were the numerical experiments on the apparent divergence in real space first observed in (\cite{Miller}) for systems of between N=8 and 32 particles, for which obviously no relaxation process has any sense. Properly performed numerical experiments (\cite{EZ}) confirmed the timescale obtained in (\cite{GS}) via Maupertuis principle.} The dynamical timescale is responsible for reaching a coarse-grained state in non-stationary systems (violent relaxation) (\cite{LB}).  

Numerical studies possess their own difficulties, starting from the choice of descriptors up to the interpretation of the results, e.g., as shown in (\cite{GK_L}), the computer image of the Lyapunov exponents is non-equivalent to their definition.      
So, the importance of further searches of strict methods to study the chaos in stellar systems is doubtless. We revisit the problem of both N-body relaxation and its time scale using both geometric (\cite{AK}) and Van Kampen stochastic differential equation approaches (\cite{VK}). The results confirm those of the Maupertuis principle for the collective relaxation timescale of collisionless spherical systems\footnote{For a study of systems with rotational momentum (spiral galaxies) with quite different dynamics, see e.g., (\cite{GK88})}.  

\section{Stochastic instability}

We consider an $N$-body system described by the Lagrangian
\begin{equation}
L(\x,\v)=\frac{1}{2}\sum_{a=1}^Nm_a|\v_a|^2-V(\x),
\end{equation}
where 
\begin{equation}
V(\x)=-G\sum_{a<b}\frac{m_am_b}{|\x_a-\x_b|}.
\end{equation}
Hereafter we use units $G=1$ and $m_a=1$, and 
$$
\x=(\x_1,\dots,\x_N),\quad  \v=(\v_1,\dots,\v_N),
$$ 
are the coordinates and velocities of stars, respectively, 
$$
\x_a=(\x_a^1,\x_a^2,\x_a^3),\quad \v_a=(\v_a^1,\v_a^2,\v_a^3),
$$
where $a=1,\dots,N$.

According to the theory of dynamical systems, the statistical properties of the systems can be studied from the behaviour of close trajectories (\cite{Arnold}). It is shown (\cite{AK}), that 
the evolution of the distance between two nearby trajectories, denoted by $\ell$, is described by a generalized Jacobi equation, which can be written in the form
\begin{equation}
  \ddot{\ell}+\B(\x,\v)\ell=0,
\end{equation}
where
\bea
\B(\x,\v)&=&-\frac{1}{3N-1}\left[\Delta V(\x)
+\frac{\nabla_{\v}^2V(\x)}{2(E-V(\x))}\right]\\
&+&\frac{3}{3N-1}\left[\frac{\Sum_{a=1}^N|F_a(\x)|^2}{2(E-V(\x))}
-\frac{\left(\Sum_{a=1}^NF_a(\x)\cdot\v_a\right)^2}{4(E-V(\x))^2}\right],
\eea
$E$ is the total energy of the system
\begin{equation}
E=\frac{1}{2}\Sum_{a=1}^N\v_a^2+V(\x),
\end{equation}
and
\begin{equation}
F_a(\x)=\mathop{\Sum_{b=1}^N}_{b\ne a}\frac{\x_b-\x_a}{|\x_b-\x_a|^3}.
\end{equation}
As in \cite{GS}, we assume that the system is collisionless\footnote{By collisionless systems, we understand, as usual, systems in which the direct impact of two stars have no role in their dynamics, and not the  neglect of gravitational encounters (scattering) of two stars. In this sense, collisionless are even the dense cores of star clusters and galaxies.}. We then have  $\Delta V(\x)=0$ and one can substitute
\begin{equation}
\frac{\v_a^i\v_b^j}{2(E-V(\x))}=\frac{1}{3N}\delta_{ab}\delta^{ij},
\end{equation}
to obtain
\begin{equation}
\label{ell}
  \ddot{\ell}+\omega(\x)\ell=0,
\end{equation}
where
\begin{equation}
\omega(\x)=\langle\B(\x,\v)\rangle
=\frac{1}{2N(E-V(\x))}\Sum_{a=1}^N|F_a(\x)|^2>0.
\end{equation}
One observes that $E-V(\x)$ is the total kinetic energy of the system 
\begin{equation}
E-V(\x)=\frac{1}{2}\Sum_{a=1}^N|\v_a|^2\sim\frac{1}{2}N\langle v^2\rangle.
\end{equation}
Thus,
\begin{equation}
 \omega(\x)\sim\frac{2}{N^2\langle v^2\rangle}\Sum_{a=1}^N|F_a(\x)|^2.
\end{equation}
For spherically symmetric systems, we can replace 
$|F_a(x)|^2$ with $|F(x)|^2$ (see \cite{Chandra}, \cite{Cohen}), 
where
\begin{equation}
F(\x)=\Sum_{a=1}^N\frac{\x_a}{|\x_a|^3}.
\end{equation}

We then replace $\omega$ with a stochastic process 
(cf. \cite{CPC},~\cite{Chandra_st}). Let $X_1,\dots, X_N$ be a sequence of $N$ independent and identically distributed (i.i.d.) random variables each having finite values of expectation $\mu$ and variance $\sigma^2 > 0$. The central limit theorem states that as $N$ increases, the sum of $N$ random variables given by
\begin{equation}
S_N = X_1 + \dots + X_N
\end{equation}
approaches the normal distribution $\mathfrak{n}(\mu,\sigma)$.
Then, at large $N$, $S_N$ will behave like a Gaussian process. This can be written as
\begin{equation}
\frac{S_N-N\mu}{\sigma\sqrt{N}}\sim\mathfrak{n}(0,1).
\end{equation}
In our case, we have $X=|F(\x)|^2$,
\begin{equation}
\mu=\langle X \rangle = \langle |F(\x)|^2 \rangle,\quad
\sigma^2=\mbox{Var}(X)=\langle|F(\x)|^4\rangle-\mu^2.
\end{equation}
Thus,
\begin{equation}
 \omega\sim\omega_0+\omega_1\mathfrak{n}(0,1),
\end{equation}
where
\begin{equation}
\omega_0\sim\frac{\mu}{N\langle v^2\rangle},\quad
\omega_1\sim\frac{\sigma}{N^{3/2}\langle v^2\rangle}.
\end{equation}

We investigate Eq.\Eq{ell} by means of a technique developed by Van Kampen (\cite{VK}) (cf. \cite{CPC}). One can derive the second moments of $\ell$ by simply rewriting Eq.\Eq{ell} in the following form
\begin{equation}
\label{Variance}
\frac{d}{dt}
\begin{pmatrix}
\langle\ell^2(t)\rangle\\
\langle\dot{\ell}^2(t)\rangle\\
\langle\ell(t)\dot{\ell}(t)\rangle
\end{pmatrix}
=\begin{pmatrix}
0 & 0 & 2\\
2\hat{\tau}\omega_1^2 & 0 & -2\omega_0\\\
-\omega_0 & 1 &0
\end{pmatrix}
\begin{pmatrix}
\langle\ell^2(t)\rangle\\
\langle\dot{\ell}^2(t)\rangle\\
\langle\ell(t)\dot{\ell}(t)\rangle
\end{pmatrix},
\end{equation}
where (\cite{CPC})
\begin{equation}
\hat{\tau}=\frac{1}{2}\cdot
\frac{\pi\sqrt{\omega_0}}{2\sqrt{\omega_0(\omega_0+\omega_1)}+\pi\omega_1}.
\end{equation}
The system given by Eq.\Eq{Variance} has a positive Lyapunov exponent $\chi$ defined by
\begin{equation}
\chi =\frac{1}{2}\left(q-\frac{4\omega_0}{3q}\right),
\end{equation}
where
\begin{equation}
q=\left(2\hat{\tau}\omega_1^2
+\sqrt{(2\hat{\tau}\omega_1^2)^2+(4\omega_0/3)^3}\right)^{1/3}.
\end{equation}

We now estimate $\mu$ and $\sigma$ to calculate $\omega_0$, $\omega_1$, and then $\hat{\tau}$ and $\chi$. Since one has (\cite{Cohen}) 
\begin{equation}
\mu=\bigg\langle|F(\x)|^2\bigg\rangle\sim n\langle v^2\rangle,
\end{equation}
where $n$ is the mean concentration of stars in the system, then 
\begin{equation}
 \sigma^2=\langle|F(\x)|^4\rangle-\mu^2\sim n^2N^2\langle v^2\rangle^2.
\end{equation}
Therefore,
\begin{equation}
 \omega_0\sim\frac{n}{N},\quad
 \omega_1\sim\frac{n}{\sqrt{N}},\quad
 \hat{\tau}\sim\frac{1}{2\sqrt{n}}\sim\tfrac{1}{2}\tau_{dyn}.
\end{equation}
And finally we derive the relaxation time
\begin{equation}
\tau_{cr}\sim\chi^{-1}\sim \tau_{dyn} N^{1/3},
\end{equation}
confirming the result of Eq.\Eq{RelTime} derived in (\cite{GS}).

\section{Conclusion}

The stochastic equation approach used above complements the probabilistic approach of Chandrasekhar and von Neumann (1943). Both are supported by the fact of decay of the time correlation function due to the exponential instability
and the Holtsmark distribution of the fluctuating force.\footnote{At distances of the order of the radius of gravitational influence 
$r_h = 1/\langle v^2\rangle$ the Holtsmark distribution diverges and should be cut off, as done in (\cite{Chandra_st}, \cite{GS}). For real stellar systems however, this cutoff is insensitive to the precise value of $r_h$, since for them $r_h$ is far smaller than the mean interstellar distance $d$ and hence the Holtsmark law vanishes far earlier.}

Thus, the stochastic equation method confirms the purely geometric derivation of the collective relaxation time given by Eq.(\ref{RelTime}). Although that formula is also supported also by alternative numerical analyses (\cite{antoni}), the present derivation avoids any approximations in numerical simulations.

Chaotic effects could be useful in constraining observationally modified gravity theories in the Solar system and galaxies (see \cite{Cap}).

\end{document}